\documentclass[12pt,preprint]{aastex}

\slugcomment{submitted to ApJ}

\shorttitle{Circumstellar Disks in Cha~I}
\shortauthors{I. Damjanov et al.}

\begin{document}

\title{A Comprehensive View of Circumstellar Disks in Chamaeleon I: \\
Infrared Excess, Accretion Signatures and Binarity} 


\author{Ivana Damjanov\altaffilmark{1}, Ray Jayawardhana\altaffilmark{1}, 
Alexander Scholz\altaffilmark{2},\\ 
Mirza Ahmic\altaffilmark{1}, Duy C. Nguyen\altaffilmark{1}, Alexis Brandeker\altaffilmark{3}, Marten H. van Kerkwijk\altaffilmark{1}}
\altaffiltext{1}{Department of Astronomy and Astrophysics, University of Toronto, 50 St.~George Street, Toronto, 
ON M5S~3H4, Canada; damjanov@astro.utoronto.ca.}
\altaffiltext{2}{School of Physics \& Astronomy, University 
of St. Andrews, North Haugh, St.Andrews, Fife KY16~9SS United Kingdom.}
\altaffiltext{3}{Stockholm Observatory,
AlbaNova University Center, Stockholms Center for Physics, Astronomy and Biotechnology,
Roslagstullsbacken 21, SE-106 91 Stockholm, Sweden.}

\email{}






\begin{abstract}
We present a comprehensive study of disks around 81 young low-mass stars and brown 
dwarfs in the nearby $\sim$2~Myr-old Chamaeleon~I star-forming region. We use mid-infrared 
photometry from the \it{Spitzer Space Telescope}\rm, supplemented by findings from ground-based 
high-resolution optical spectroscopy and adaptive optics imaging. We derive disk fractions 
of $52\%\pm6\%$ and $58^{+6}_{-7}\%$ based on 8~$\mu$m and 24~$\mu$m colour excesses, respectively, 
consistent with those reported for other clusters of similar age. Within the uncertainties, 
the disk frequency in our sample of K3--M8 objects in Cha~I does not depend on stellar mass. 
Diskless and disk-bearing objects have similar spatial distributions. There are no obvious 
transition disks in our sample, implying a rapid timescale for the inner disk clearing 
process; however, we find two objects with weak excess at 3--8~$\mu$m and substantial excess at 
24~$\mu$m, which may indicate grain growth and dust settling in the inner disk. For a 
sub-sample of 35 objects with high-resolution spectra, we investigate the connection 
between accretion signatures and dusty disks: in the vast majority of cases (29/35) 
the two are well correlated, suggesting that, on average, the timescale for gas 
dissipation is similar to that for clearing the inner dust disk. The exceptions are 
six objects for which dust disks appear to persist even though accretion has 
ceased or dropped below measurable levels. Adaptive optics images of 65 of our targets reveal 
that 17 have companions at (projected) separations of 10--80~AU. Of the five 
$\lesssim20$~AU binaries, four lack infrared excess, possibly indicating that a close 
companion leads to faster disk dispersal. The closest binary with excess is separated 
by $\sim20$~AU, which sets an upper limit of $\sim8$~AU for the outer disk radius. The 
overall disk frequency among stars with companions ($35^{+15}_{-13}\%$) is lower than 
(but still statistically consistent with) the value for the total sample.

\end{abstract}


\section{Introduction}
Circumstellar disks are a natural outcome of star formation and the
potential birth sites of planets. Thus studies of disk evolution can
provide useful constraints on both pre-main-sequence stellar
evolution and planet formation timescales. Dusty disks may be
identified most easily by their infrared emission. Observations in
the mid-infrared are much more sensitive to disks than those in
the near-infrared, since the stellar photospheric flux declines rapidly at
longer wavelengths while the disk emission rises. The {\it Spitzer Space
Telescope}, with its Infrared Array Camera (IRAC) and Multiband
Imaging Photometer (MIPS) operating in the 3.6--160~$\mu$m range, has
greatly improved our ability to identify  and study disks surrounding
young stars (e.g., \citealp{Hartmann2005,sa2006,Lada2006}). Processes
of grain growth and dust settling to the midplane, along with the
possible formation of planetesimals, are thought to occur 
in the inner disks. The consequent reduction of
small dust grain abundance results in weak or absent  near-infrared 
emission but the mid- to far-infrared disk emission remains strong,
leading to the characteristic signature of ``transition disks'' 
\citep{D'Alessio2005,calvet2005,Muzerolle2006}. The spectral energy 
distribution (SED) in 4--25~$\mu$m wavelength range is thus ideal for 
probing dust in the inner disk. 

Since gas constitutes most ($\sim99\%$) of the disk mass and is 
essential for building giant planets, it is important to know whether 
gas and dust disperse on similar timescales. While it is difficult 
to detect the bulk of the disk gas directly, its presence can be 
inferred from signatures of accretion from the inner disk edge onto 
the central star. The most prominent signature of
infalling gas is a broad, asymmetric and strong H$\alpha$ emission
line \citep{Mohanty2005,Muzerolle2005,Jayawardhana2006}. A comparison
between accretion signatures and infrared emission can reveal whether
the timescales for dust and gas dissipation are similar. Another
important question related to planet formation is whether the presence
of a binary companion has a considerable effect on disk evolution
\citep[and references therein]{Bouwman2006}.

Here we present a comprehensive study of disks surrounding 81 stars and
brown dwarfs in the nearby ($160$~pc; \citealp{Whittet1997}), young ($\sim$2
~Myr; \citealp[and references therein]{Luhman2004}) Chamaeleon~I cloud (abbreviated Cha~I in the following),
using \t{Spitzer} \rm  IRAC 4--8~$\mu$m and MIPS 24~$\mu$m data. We study the
overall disk frequency as well as its possible dependence on stellar
spectral type, and look for transition disk candidates. For 35 of the
targets, for which high-resolution optical spectra are available (Nguyen
et al., in prep.; \citealp{Mohanty2005,Muzerolle2005}), we investigate the
connection between infrared excess and accretion signatures. For 65
objects --the largest such sample to date-- the availability of
adaptive optics images (Ahmic et al., in prep.) allows us to explore how
multiplicity affects disk dissipation.

\section{Spitzer Data and Photometry}\label{data}

We used publicly available images of the Cha~I star-forming region from 
the IRAC and MIPS instruments aboard the \it {Spitzer Space Telescope} \rm (GTO programs \# 6 and 37, PI: G. Fazio). 
The IRAC images of the southern and northern cloud cores in Cha~I at 3.6, 4.5, 5.8 and 8.0~$\mu$m were obtained in 
June and July 2004 (AOR keys 3960320 and 3651328). The plate scale and the spatial resolution of IRAC are $1\farcs2$ 
and $\sim2\arcsec$, respectively \citep{Fazio2004}. For these programs, the Spitzer Science Center (SSC) pipeline 
(ver S14.0.0) provides two shallow (exp.~time 20.8~s) maps for all of the four IRAC channels, each performed twice 
with an offset of a few arcseconds. These dual maps for the southern cloud have centers at 
$\alpha\approx11^\mathrm{h}07^\mathrm{m}$, $\delta\approx-77^\mathrm{h}34^\mathrm{m}$ 
(J2000) for 3.6 and 5.8~$\mu$m and  at $\alpha\approx11^\mathrm{h}07^\mathrm{m}$, 
$\delta\approx-77^\mathrm{h}28^\mathrm{m}$ (J2000) for 4.5 and 8.0~$\mu$m. 
The northern cloud core maps have centers 
at $\alpha\approx11^\mathrm{h}09^\mathrm{m}$, $\delta\approx-76^\mathrm{h}36^\mathrm{m}$ 
(J2000) for 3.6 and 5.8~$\mu$m and at $\alpha\approx11^\mathrm{h}10^\mathrm{m}$, 
$\delta\approx-76^\mathrm{h}30^\mathrm{m}$ (J2000) for 4.5 and 8.0~$\mu$m. 
The dimensions of the maps are $33\arcmin\times29\arcmin$. In each of the four bands, BCD images for each of the 
shallow maps were taken through the post-BCD pipeline at the SSC, where the pointing refinement and mosaicing with 
standard MOPEX tool were performed. We used these archived final post-BCD shallow mosaics for IRAC photometric 
measurements in our analysis.

The observations of the northern and southern Cha~I cloud cores using MIPS were taken in April 2004 and February 2005 
(AOR keys 3661312 and 3962112). The relevant MIPS imaging band for the purpose of this analysis is the one centered at 
24~$\mu$m, since the sensitivity limits and background emission in MIPS channels at 70~$\mu$m and 160~$\mu$m do not allow 
detection of any but the brightest mid-infrared sources. The plate scale of the 24~$\mu$m array is $2\farcs45$; it has a 
spatial resolution of 6$\arcsec$ \citep{Rieke2004}. The Cha~I maps have a total size of about $0\fdg5\times1\fdg5$. The 
total effective exposure time per pointing at 24~$\mu$m is about 80~s. After calibration and correction for distortion, 
images were combined into the final mosaics within the standard post-BCD SSC pipeline. We used these archived MIPS mosaics
for 24~$\mu$m flux measurements.

\begin{figure}[htp]
\begin{center}
\includegraphics[scale=0.8]{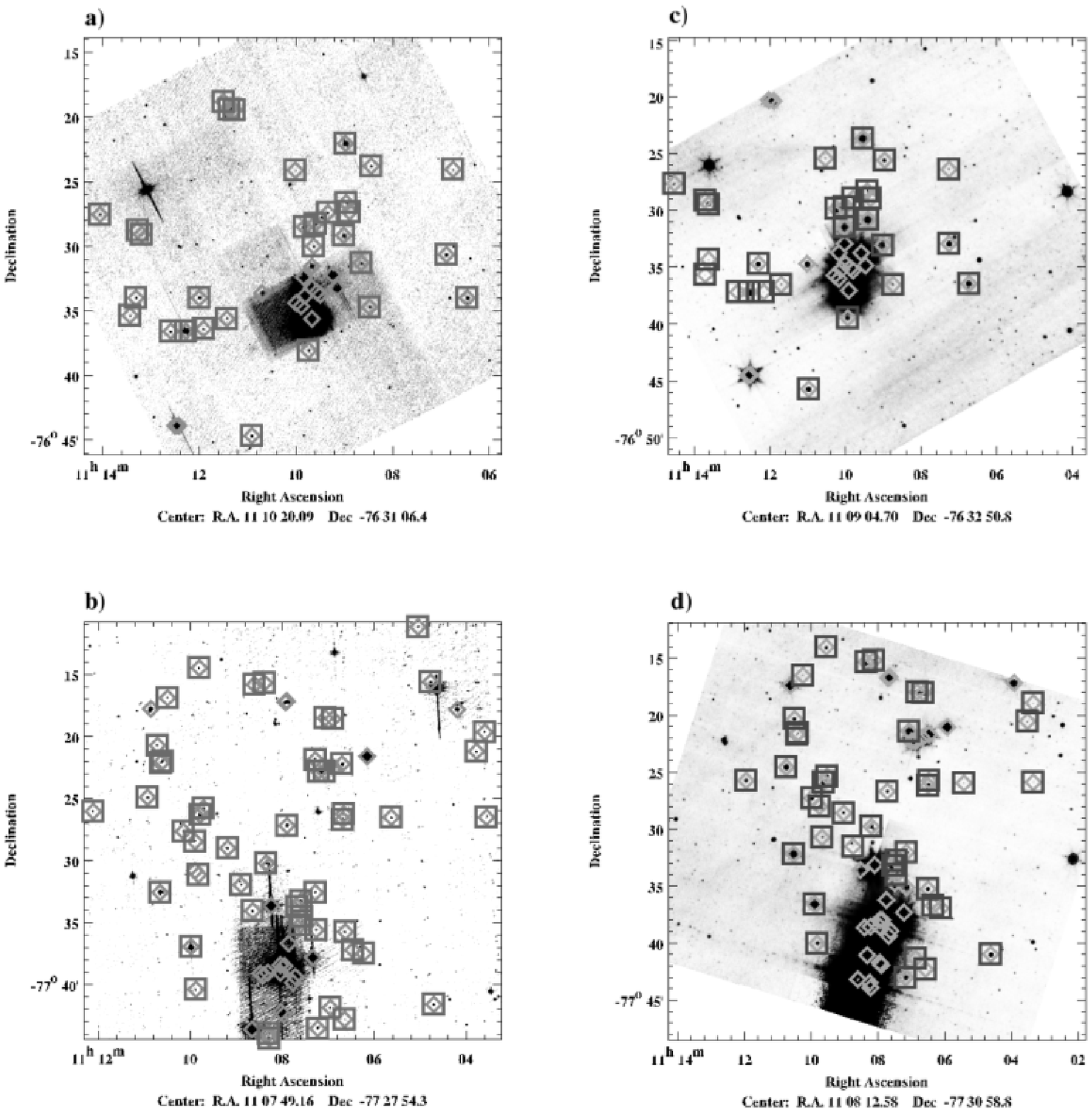}
\caption{IRAC 8~$\mu$m and MIPS 24~$\mu$m mosaics with the positions of all detected Cha~I members denoted with \it{diamonds} \rm and positions of the
objects used in our analysis presented with \it{squares}\rm; a) upper cloud core mosaic in  8~$\mu$m band, b) lower cloud core mosaic in  8~$\mu$m band,
c)  upper cloud core mosaic in  24~$\mu$m band, d) lower cloud core mosaic in  24~$\mu$m band. 
\label{f1}}
\end{center}
\end{figure}

We considered the 157 Cha~I members listed in \citet{Luhman2004}. These are mostly late-type stars with 
spectral classes in the range of K0--M8, including eight likely brown dwarfs with spectral types later than M6. 
More than 100 members of Cha~I are found in the IRAC and MIPS images used in our analysis. Two small areas in 
the covered regions, one each for the northern and the southern field, are affected by localized extended emission 
associated with bright stars, overlapping PSF structures, as well as saturation effects (see Fig.~\ref{f1}). For objects
in those areas, aperture photometry does not give reliable results. To be conservative, we excluded such objects from 
our analysis -- 35 out of 116 in IRAC and 39 out of 126 in MIPS. We note that this excludes all stars brighter than K3. 
Our final list of targets contains 81~Cha~I objects with spectral types that span from K3 to M8. Out of these, 69 have 
detectable fluxes in all four IRAC channels and in the MIPS 24~$\mu$m band. Additional 18 objects with measured 
24~$\mu$m fluxes were not present in the IRAC field of view. IRAC 8~$\mu$m and MIPS 24~$\mu$m mosaics with all 
detected objects are presented in Fig.~\ref{f1}. 

We performed photometry using the aperture photometry routine \texttt{aper.pro} in the IDLPHOT package. 
The IDLPHOT routines were integrated into a custom IDL program that uses the World Coordinate System information in 
the image headers to find the star within the image. The program then determines the centroid of the star and extracts 
the photometry. For IRAC data, we used a 5-pixel ($6\farcs1$) aperture for each object, with a sky annulus extending 
from 5 to 10 pixels for sky flux measurements. We applied an aperture correction following the instructions given in the 
IRAC handbook, version 3.0. For eight Cha~I brown dwarfs with IRAC fluxes previously published by \citet{Luhman2005}, 
our photometry is consistent with theirs to within the errors. 
For MIPS 24~$\mu$m images, we chose a $6\farcs6$ aperture, with a sky annulus in the range 
of 7-13$\arcsec$. In order to account for the aperture correction, we used values tabulated in the MIPS handbook, version 
3.2.1. The final list of Cha~I targets with their fluxes in units of mJy measured in the 3.6, 4.5, 5.8, 8~$\mu$m and 24~$\mu$m 
bands is given in Table~I. The $J-H$ colour for all Cha~I members listed in Table~I is from the Two Micron All Sky 
Survey (2MASS) Point Source Catalog.  

\section{Disk Frequency}\label{frac}

\begin{figure}[htp]
\begin{center}
\includegraphics[scale=0.8]{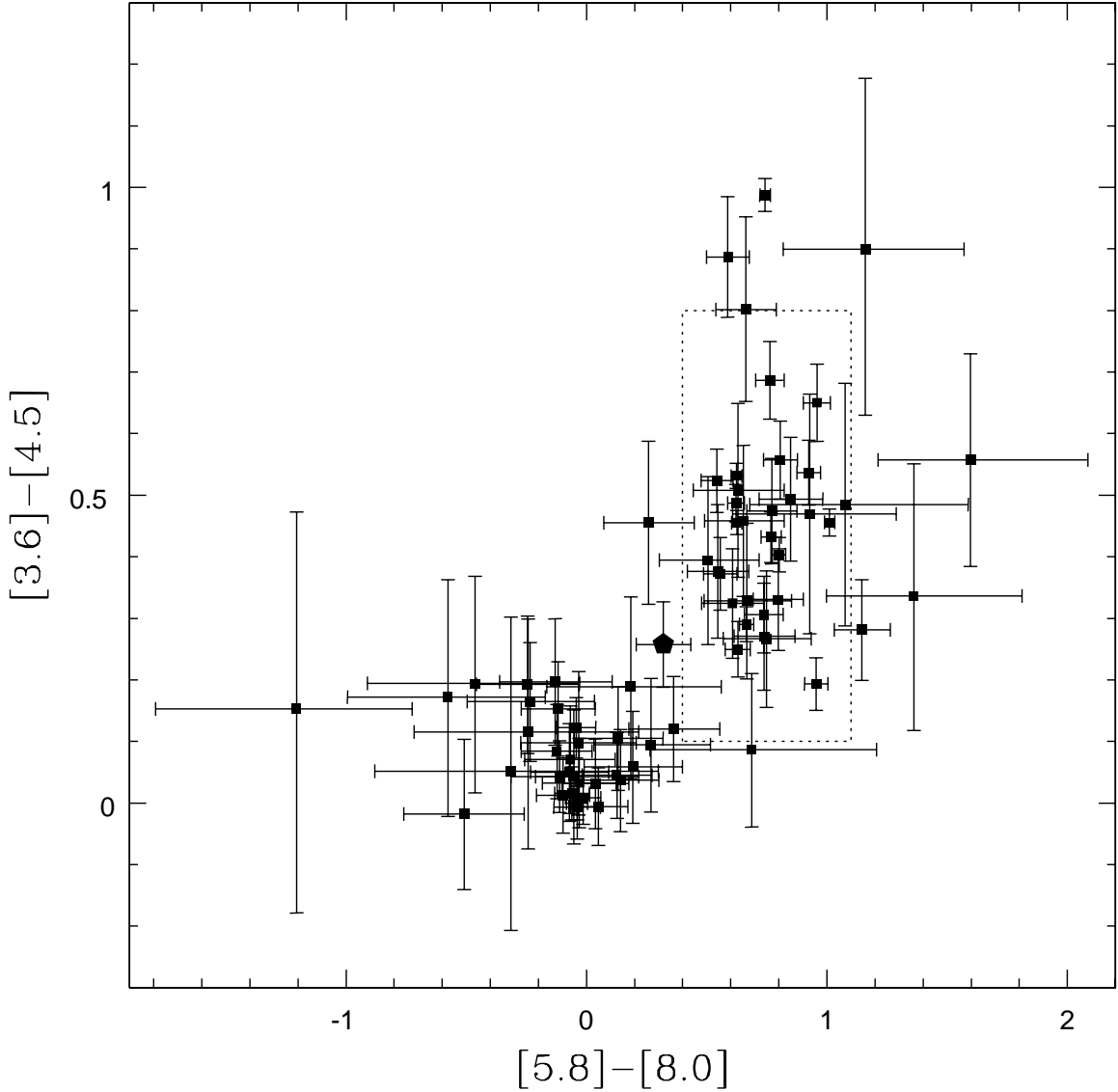}
\caption{IRAC colour-colour diagram for 71 objects in Cha~I star-forming region. The \it{dotted square} \rm denotes 
the domain of objects with circumstellar disks \citep{allen2004,Hartmann2005}. Objects without disk excess 
are located near the origin of the coordinate system. One transition object candidate --C7-1-- with measured fluxes in all four IRAC 
channels is presented with \it{pentagon}\rm.
\label{f2}}
\end{center}
\end{figure}

It is possible to distinguish between diskless (so-called Class III) and disk-harboring (Class II) stars using 
the IRAC colours, as suggested by \citet{allen2004}. The two classes of objects are indeed well separated in 
our IRAC colour-colour diagram (Fig.~\ref{f2}): diskless stars do not show colour excess in the IRAC channels 
and thus fall near the origin of the coordinate system, while those with disks are concentrated in a region demarcated 
by model colours and measured IRAC colours for objects in various other star-forming regions \citep{allen2004,Hartmann2005}. 

\begin{figure}[htp]
\begin{centering}
\includegraphics[scale=.43]{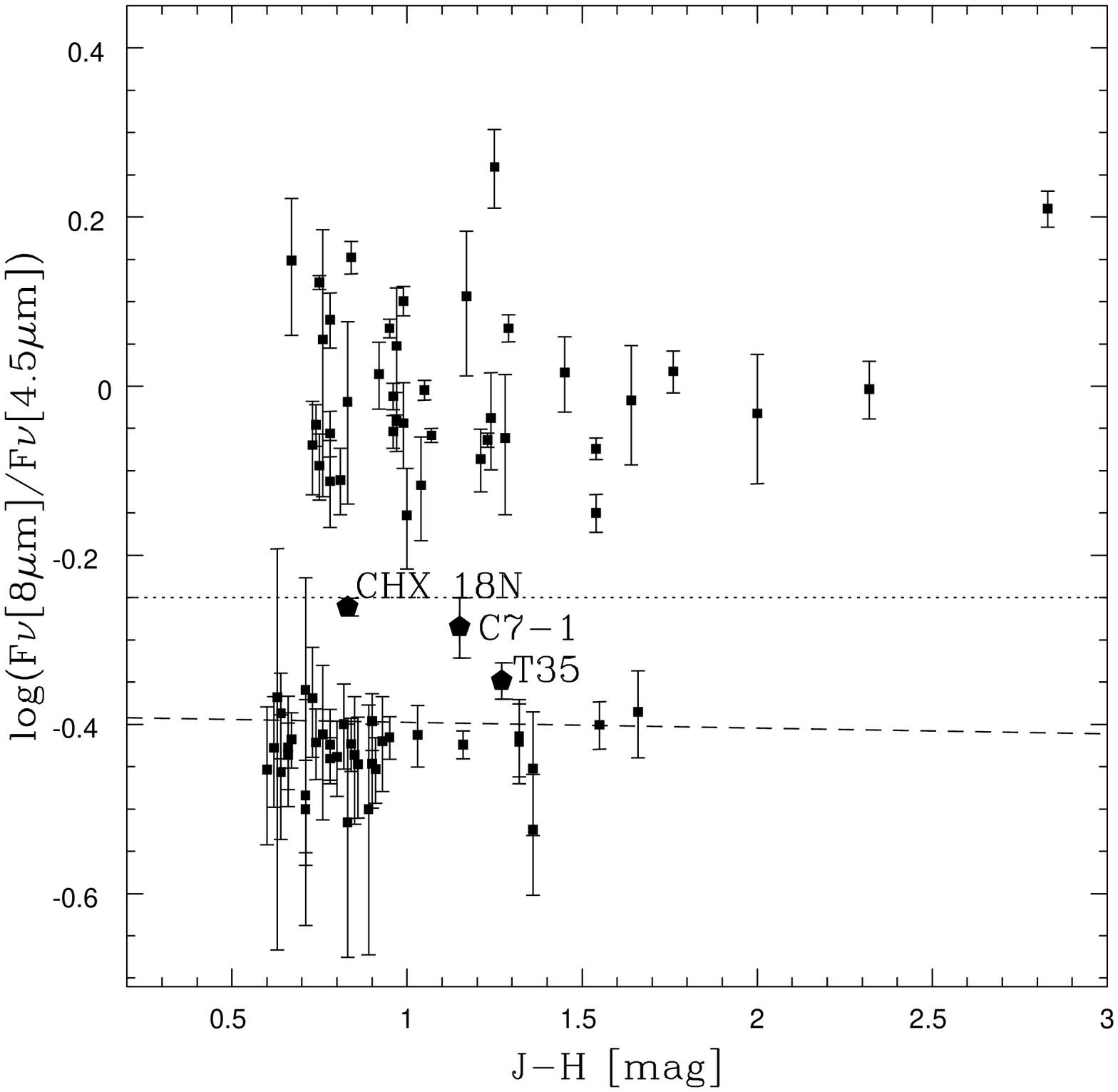}
\hfill
\includegraphics[scale=.43]{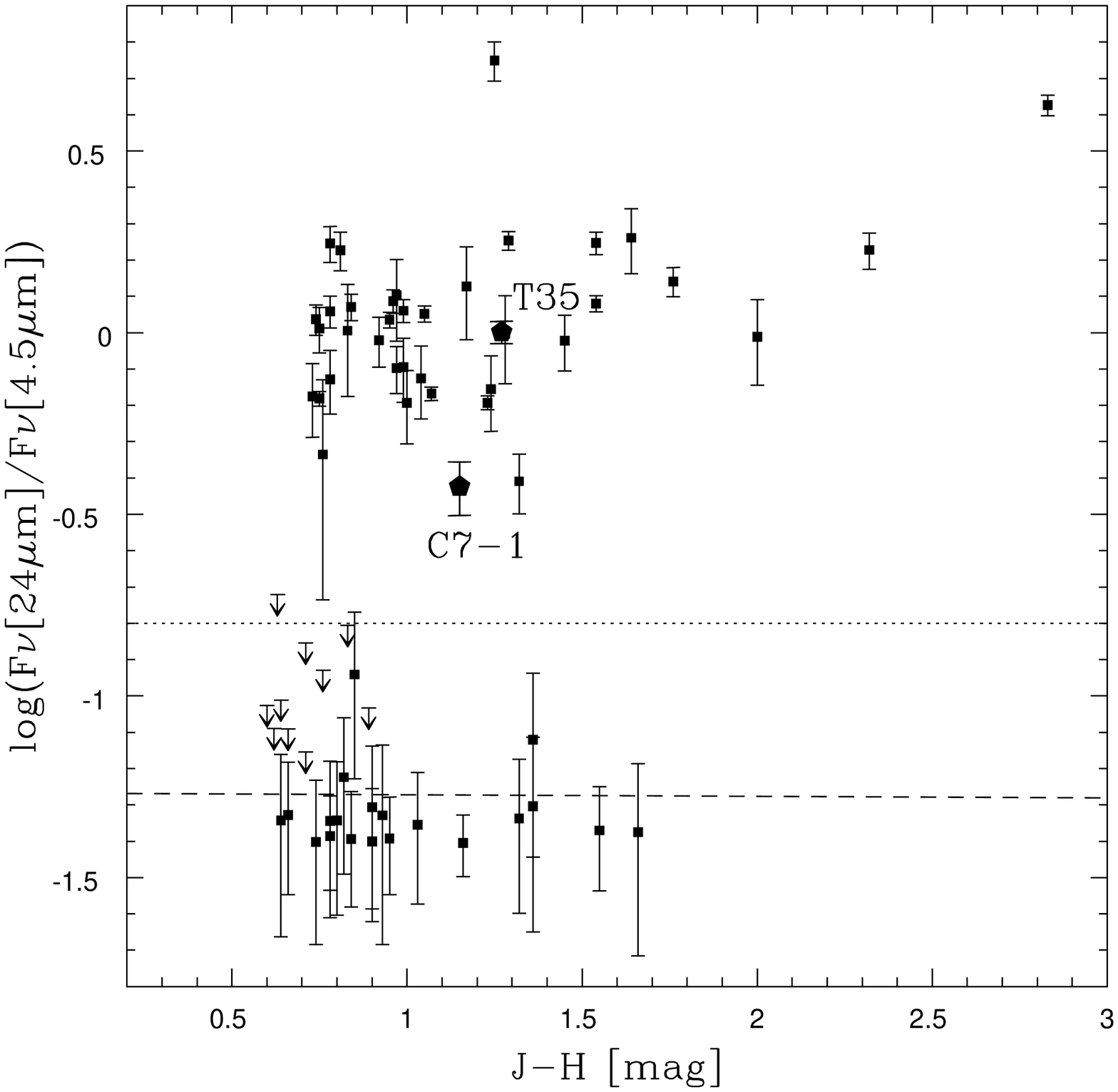}
\caption{Colour-colour diagrams with circumstellar disks tracers (8~$\mu$m to 4.5~$\mu$m flux ratio in the top panel and 
24~$\mu$m to 4.5~$\mu$m ratio in the bottom panel) vs. a stellar photosphere tracer ($J-H$). Arrows in the lower panel 
represent objects for which only upper limits for 24~$\mu$m fluxes could be estimated. \it{Dashed line} \rm in 
both panels corresponds to the photospheric 8~$\mu$m to 4.5~$\mu$m flux ratio. The \it {dotted lines} \rm represent 
conservative thresholds for circumstellar disk presence, in our case a 40\% excess above the photosphere in the 
8/4.5~$\mu$m and 24/4.5~$\mu$m flux ratios. 
There are two objects that appear to lie under the threshold in the top panel, but occupy the region above the threshold 
in the bottom one. These are candidates for the transition objects, i.e. stars with possible clearings in their disks. 
A third object from the upper panel --CHX~18N-- does not have 24~$\mu$m flux measured. All three object are presented with 
\it {pentagons} \rm and labeled.  
\label{f3}}
\end{centering}
\end{figure}

Another way to distinguish stars with disks from those without is to plot mid-infrared flux ratios against 
the $J-H$ colour, which is a reasonable proxy for the stellar photosphere. We do this in Fig.~\ref{f3} 
for the 8~$\mu$m to 4.5~$\mu$m and 24~$\mu$m to 4.5~$\mu$m flux ratios. The presence of two populations of stars 
is evident. The population with lower flux ratios shows the infrared emission dominated 
by the stellar photospheres: these are comparable to the blackbody emission function with the temperature range 
that includes the effective temperatures of all Cha~I members in our sample (dashed line in both panels of 
Fig.~\ref{f3}). Objects with the higher flux ratios are the ones likely to harbor circumstellar disks; dust 
emission from the surrounding disk contributes substantially to the infrared emission coming from these objects. 
Interestingly, there is a clear gap between diskless and disk-bearing stars in both panels, indicating a rapid 
transition between these two phases (see~\S~\ref{trans} for further discussion). 

Based on 8~$\mu$m excess, the disk fraction among Cha~I members included in our analysis is $52\%\pm6\%$ ($42/81$) 
while for objects detected at 24~$\mu$m, it is $58^{+6}_{-7}\%$ ($40/69$)\footnote{Cautious note: Since we excluded 
a significant fraction of the Cha~I members located close to bright stars (see \S~\ref{data}), our disk frequencies may be 
biased.}.
 Thus, the disk frequency we derive for Cha~I is comparable 
to that found in the $\sim$2~Myr-old IC~348 cluster ($50\%\pm6\%$) from IRAC observations \citep{Lada2006}, but somewhat lower 
than the $\sim68\%$ disk fraction reported for the $\sim$1~Myr-old Taurus star-forming region, based on 3.6~$\mu$m excess 
measurements \citep{Haish2001}. On the older side, only $19\%\pm4\%$ of K+M stars in the $\sim$5 ~Myr-old Upper Scorpius OB 
association show disk excess at 8~$\mu$m and 16~$\mu$m \citep{Carpenter2006}. A similar steep decline in the frequencies 
of the low-mass stars with disks is also seen in the $\sim$8 ~Myr-old TW Hydrae association \citep{Jayawardhana1999} and 
in the $\sim$10--12~Myr old NGC~7160 cluster \citep{sa2006}.

\begin{figure}[htp]
\begin{center}
\includegraphics[scale=0.8]{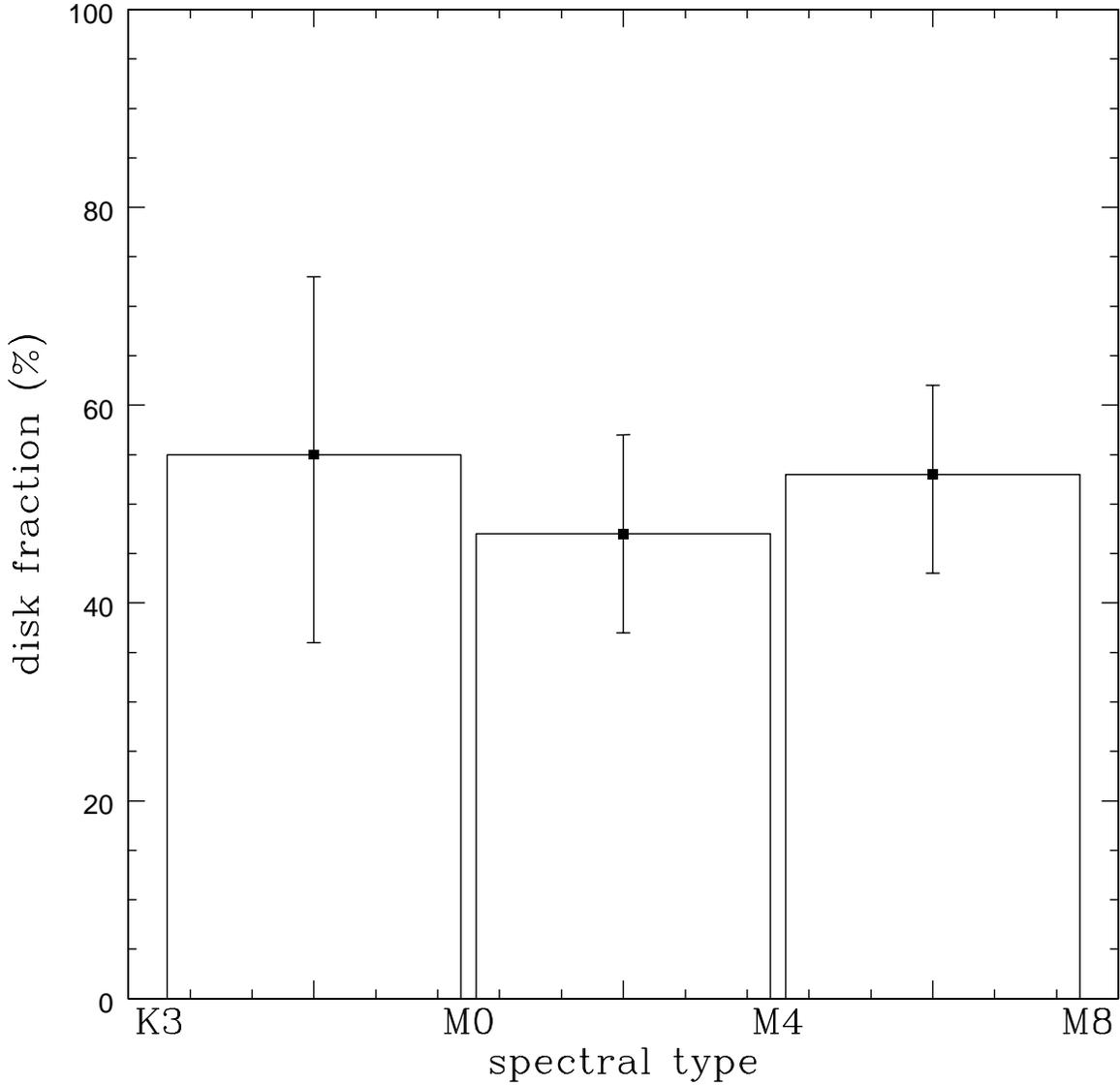}
\caption{Histogram of the disk fraction as a function of spectral type for Cha~I members. The spectral type information is 
taken from \citet{Luhman2004} while the disk detection is based on the 8~$\mu$m excess. The majority of stars in Cha~I and all 
objects included in this study are of spectral types later than K3. In order to have roughly the same number of objects in every bin, 
we divided our sample into three bins. The obtained distribution shows that there is no significant difference between the disk 
fractions among stars of different (late) spectral types in Cha~I.
\label{f4}}
\end{center}
\end{figure}

There have been suggestions that the disk fraction among young stars depends on spectral type and stellar mass 
\citep{Lada2006,Carpenter2006,sa2006}. However, in the case of Cha~I, circumstellar disks appear to be distributed evenly 
among late-type ($\geq$~K3) stars (Fig.~\ref{f4}). The fraction of disk-bearing stars in our sample ranges from 
$47\%\pm10\%$ for M0--M4 stars to $53^{+9}_{-10}\%$ for M4.25--M8 stars and $55^{+18}_{-19}\%$ for K3--K8 stars. These 
disk fractions are similar to that ($50\%\pm17\%$) reported by \citet{Luhman2005} for very low mass objects in this region. 
In order to consider a possible dependance on stellar mass, we can convert the effective temperatures of Cha~I members in 
our sample \citep{Luhman2004} to masses using the \citet{Baraffe1998} pre-main-sequence (PMS) evolutionary tracks for an 
age of $\sim2$~Myr. We find that stars with masses $\gtrsim1$~M$_{\sun}$ have a disk fraction of $55^{+18}_{-19}\%$ 
while $47\%\pm10\%$ of those in the 0.3~M$_{\sun}$--$\lesssim1$~M$_{\sun}$ range harbor dusty disks. Again, we do 
not find a significant mass-dependance in the disk frequency, for Cha~I members with masses of 0.1--1.4~M$_{\sun}$. On the 
other hand, \citet {Lada2006} have found the highest percentage of circumstellar disks in the IC~348 cluster to be 
among K6--M2 stars ($\sim56\%$), which have masses comparable to the Sun, while the disk fraction for stars $<$ K6 
is only $\sim19\%$. For Upper Scorpius, the highest disk fraction is for K and M stars with masses in the 0.1--1~M$_{\sun}$ 
range \citep{Carpenter2006}, although the peak value ($\sim19\%$) is much lower than in the younger Cha~I and IC~348 regions. 
Meanwhile, $\sim48\%$ of G--M2 stars in Trumpler~37 ($\sim$4~Myr) show infrared excess \citep{sa2006}. In summary, the 
evidence to date suggests that primordial disks survive for $\sim$5 ~Myr around 20--50\% of the stars with masses 
$\lesssim1$~M$_{\sun}$, thus providing conditions for planet formation. The disk lifetime seems to be a function 
of object mass in some regions, but not in Cha~I, thus it may also depend on the environment.

\begin{figure}[htp]
\begin{center}
\includegraphics[scale=0.8]{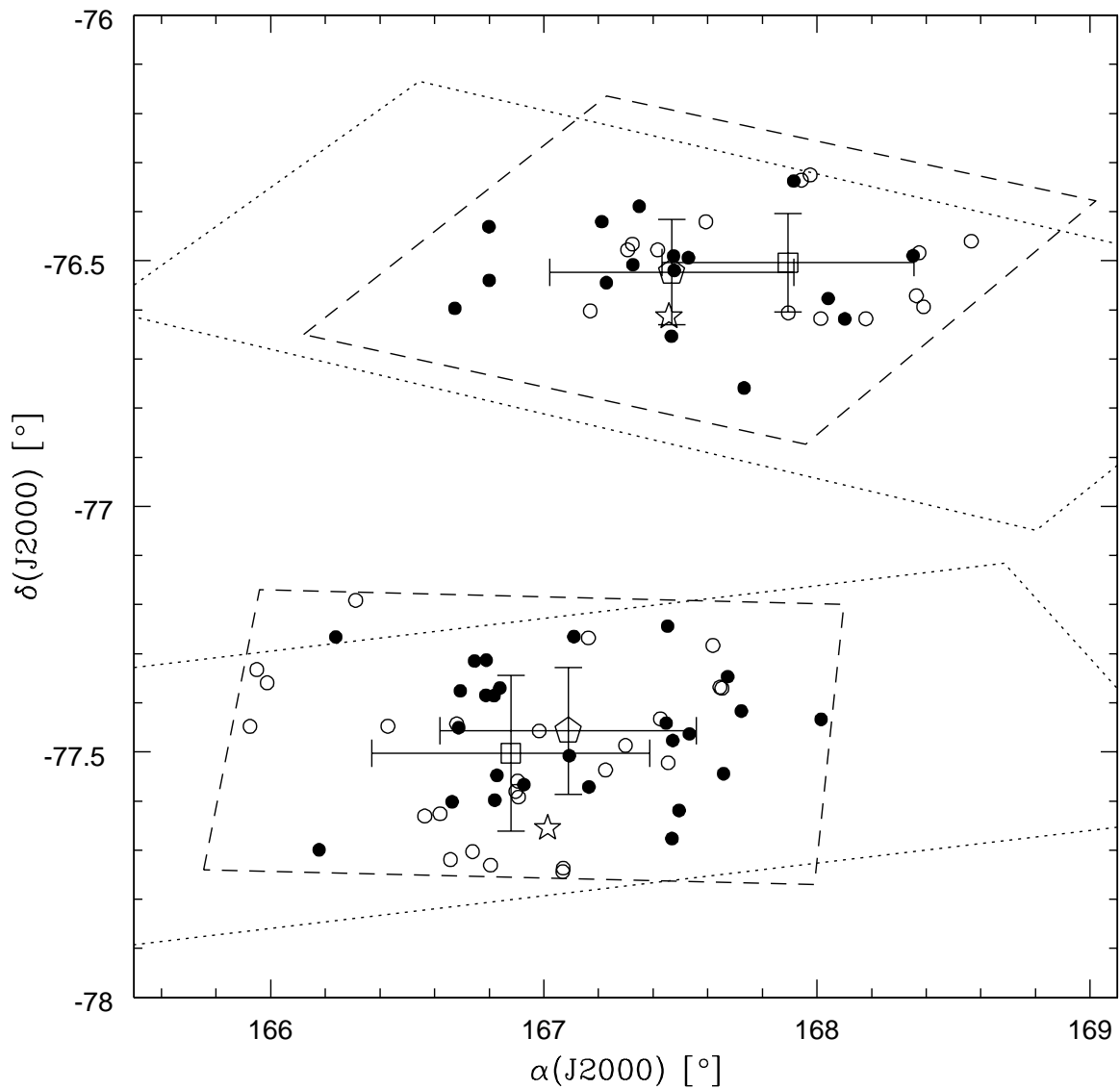}
\caption{Spatial distribution of Cha~I members: \it {filled circles} \rm correspond to the positions of the disk bearing objects, 
while objects without any infrared excess due to disk presence are denoted with \it {open circles}\rm. The two \it{stars} \rm represent 
intermediate mass stars - HD~97048 in the lower cloud core and HD~97300 in the upper one. \it{Open squares} \rm and \it{open pentagons} \rm 
mark the average positions of diskless and disk-bearing stars, respectively, in the two cores. \it{Dashed lines} \rm correspond to the boundaries of 
IRAC maps, and \it {dotted lines} \rm correspond to the boundaries of the 24~$\mu$m MIPS maps.   
\label{f5}}
\end{center}
\end{figure}

We also investigated whether there is a difference between the spatial distribution of diskless stars versus disk-bearing 
stars, especially in terms of their apparent distance from the two intermediate-mass stars in Cha~I. The Herbig Ae/Be star 
HD~97048 is located in the lower cloud core, while the likely zero-age main sequence A star HD~97300 is in the upper cloud 
core. Figure~\ref{f5} shows the positions of our targets along with those of the two early-type stars. There does not appear 
to be a difference between the spatial distributions of stars with 8~$\mu$m excess and those without (filled and open circles in 
Fig.~\ref{f5}, respectively), outside of $2.5\arcmin$ from HD~97300 in the upper cloud and $6\arcmin$ from HD~97048 
in the lower cloud core; at smaller distances contamination from the bright stars and cloud core emission prevent reliable 
flux measurements (see \S~\ref{data}). Average positions of the two types of objects in the lower and upper cloud regions, denoted 
as open pentagons for disk-bearing and open squares for diskless stars, might suggest that disk-bearing objects are slightly 
displaced towards the central region of Cha~I. However, this displacement is too small to be statistically significant. Thus, we 
conclude that there is no evidence for intermediate mass stars influencing disk dissipation in low-mass stars in Cha~I.
   
\section{Transition Disks}\label{trans}
Objects that exhibit fluxes at 3--8~$\mu$m fully consistent with photospheric emission, but higher flux levels at 
longer wavelengths are thought to be caught in the process of inner disk clearing. These disks may have an optically 
thin inner region devoid of small dust grains, as in the case of CoKu Tau/4 in Taurus \citep{D'Alessio2005}. Theoretical 
models proposed to explain the clearing process include photoevaporation by ultraviolet radiation from the central star 
and grain growth related to planet formation processes \citep{Muzerolle2006}.    

\begin{figure}[htp]
\begin{center}
\includegraphics[scale=0.8]{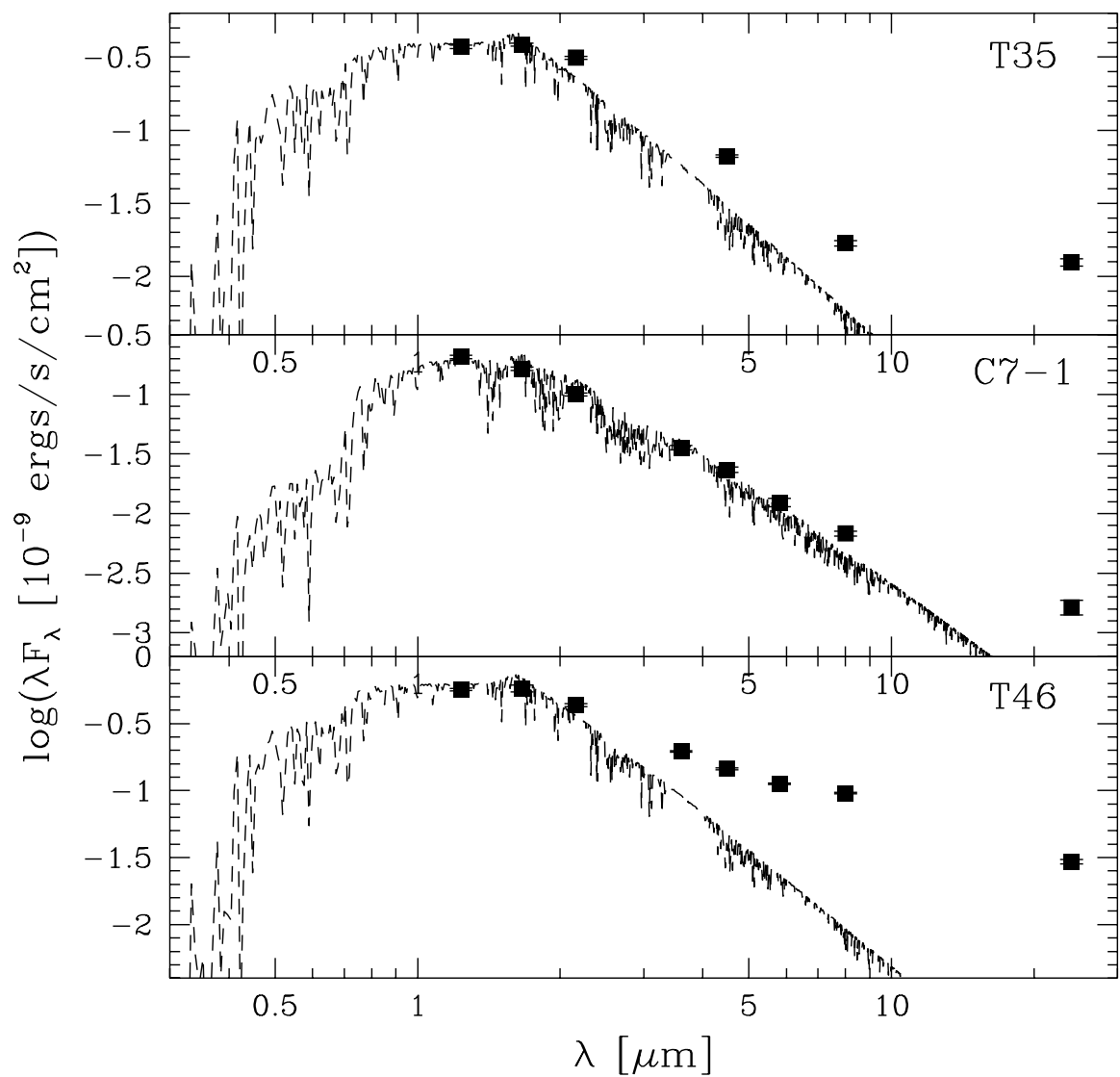}
\caption{SEDs for two transition object candidates (T35 and C7-1) and one typical disk-bearing Cha~I member (T46). In each panel measured fluxes in 
2MASS, IRAC and 24~$\mu$m MIPS bands are presented with \it{squares}\rm; the model stellar photosphere corresponding to the object's effective temperature 
(STARdusty2000, \citealt{Allard2000}) is shown with the \it{dashed line} \rm for comparison.
\label{f6}}
\end{center}
\end{figure}

To identify transition disk candidates among Cha~I members in our sample, we have searched for objects with no 
infrared colour excess at 3--8~$\mu$m (i.e., lying  under the dotted line in the upper panel of Fig.~\ref{f3}), 
but with excess flux at 24~$\mu$m (i.e.,  lying   above  the  dotted  line in  the lower panel of  Fig.~\ref{f3}). 
We found two candidates: K8-type star T35 and M5-class star C7-1. We then plotted the infrared spectral energy 
distributions (SEDs) of these two objects, including near-infrared fluxes from the 2MASS Point Source Catalog, 
along with the SED of the disk-bearing star T46 (K8) for comparison (Fig.~\ref{f6}). All fluxes have been 
dereddened using $A_{J}$ given in \citet{Luhman2004} and the reddening law of \citet{Mathis1990}. All three objects show 
excess above the expected photospheric emission at $\lesssim10$~$\mu$m (based on the STARdusty 2000 models of stellar atmospheres 
by \citealt{Allard2000}), but the excess is less pronounced in 
T35 and C7-1. Since small ($<10$~$\mu$m), warm ($T>200$~K) dust grains dominate emission in this wavelength region,
the flatter SEDs of T35 and C7-1 may imply that their inner disks are undergoing grain growth and dust settling to
the optically thick midplane \citep{Scholz2007}. However, their inner disks are not completely devoid of small grains,
unlike the case in previously studied transition objects (e.g., \citealp{D'Alessio2005, Muzerolle2006}).   

\section{Connections with Accretion and Binarity}\label{other}
Our Cha~I sample is ideally suited for investigating the effects of stellar properties and immediate environment on 
disk dissipation, because high-resolution optical spectra and high-angular resolution images are available for a 
substantial fraction of the targets. The optical spectra, obtained with MIKE on the Magellan/Clay 6.5\,m telescope, 
include tracers of disk accretion for 35 objects (Nguyen et al., in prep.; \citealp{Mohanty2005,Muzerolle2005}). Adaptive optics (AO) 
imaging with NACO on the ESO Very Large Telescope has been carried out for 65 of our targets (Ahmic et al., in prep.). 

\subsection{Accretion}\label{acc}
The strength of the H$\alpha$ emission line is widely used as a signature of (gas) accretion from a disk onto the 
central star \citep[e.g.,][]{Jayawardhana2006}. In the spectra of accreting objects, the line is dominated by emission produced 
in the infalling gas, whereas in non-accretors the line is weaker because it comes only from chromospheric activity. 
Traditionally, young stars were classified as either classical T~Tauri stars (CTTS) if the H$\alpha$ equivalent width 
(EW) is $> 10$~\AA~ or as weak-line T~Tauri stars (WTTS) if the EW~$< 10$~\AA. But, since the EW depends on the spectral 
type, \citet{wb2003} proposed the full width of the line at 10$\%$ of the peak as a more robust accretion diagnostic. 
Based on empirical and physical arguments, \citet{Jayawardhana2003} adopted an H$\alpha$ $10\%$ width 
of~$\sim200$~km~s$^{-1}$ as a dividing line between accretors and non-accretors, to be applicable to the very low mass 
 regime as well. In fact, \citet{Natta2004} have found that the H$\alpha$ $10\%$ correlates well with the mass 
accretion rate derived in other ways, and that the $\sim200$~km~s$^{-1}$ threshold corresponds to the mass acretion rate
$\sim10^{-11}$~M$_{\sun}$~yr$^{-1}$. 

\begin{figure}[htp]
\begin{center}
\includegraphics[scale=0.8]{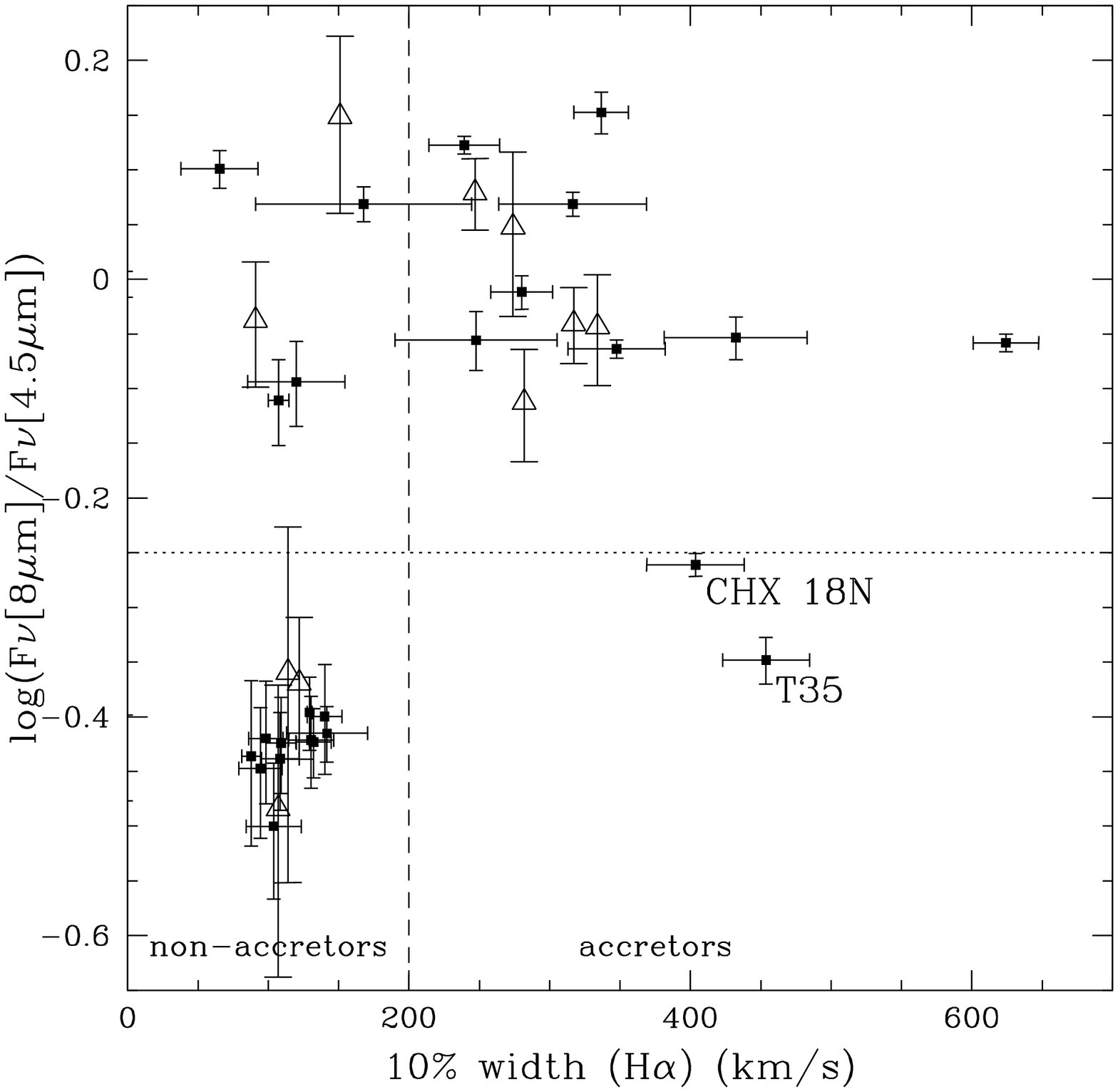}
\caption{8~$\mu$m excess vs. the full width of H$\alpha$ line at 10\% of the peak for 35 Cha~I members. 
The H$\alpha$ widths are taken from Nguyen et al. (in prep.; \it{filled squares}\rm), \citet{Mohanty2005} 
and \citet{Muzerolle2005} (\it{open triangles}\rm).
The \it{dashed line} \rm corresponds to the accretion cutoff adopted by \citet{Jayawardhana2003}. The threshold for 
disk presence on the basic of IRAC colour excess is represented by \it{dotted line} \rm as in the upper panel of 
Fig.~\ref{f3}. The two objects with prominent accretion signatures and yet without colour excess at $8/4.5\,\mu$m 
are labeled. 
\label{f7}}
\end{center}
\end{figure}

To examine the connection between dusty inner disks and gas accretion, in Fig.~\ref{f7} we show the 8~$\mu$m 
excess versus H$\alpha$~$10\%$ widths for those Cha~I members for which both measurements are available. (The error 
bars on the H$\alpha$~$10\%$ width refer to the standard deviations of the estimates over four epochs of observations.) 
Out of 35 objects in this sub-sample, 14 do not show evidence of accretion or dust emission (cf. lower left region in 
Fig.~\ref{f7}), implying that their inner disks have probably been cleared of both dust and gas. Of 21 objects 
with infrared excess, 15 appear to be accreting.\footnote{For two of the objects --CHX~18N and T35-- the 8~$\mu$m excess 
is relatively weak; the latter does have substantial excess at 24~$\mu$m (see~\S~\ref{trans}) while for the former we could not 
measure it reliably due to background contamination.} Thus, in the vast majority of cases (29/35) the signature of 
accretion is well correlated with evidence of a dusty disk. This finding suggests that, on average, the gas dissipation 
timescale is comparable to the timescale for clearing the inner disk of small dust grains. 

However, six out of the 21 disk-bearing objects in this sub-sample lie below the accretion threshold (upper left 
part of Fig.~\ref{f7}). Five out of these six objects are in the low-mass regime, with the masses $\lesssim0.3$~M$_{\sun}$ 
(see Table~\ref{tbl-2}). One possible explanation is that accretion rates have dropped below measurable levels 
even though dust disks persist in these cases. Another explanation is variable accretion: it may well be the case 
that quiescent phases dominate in the latest stages of accretion, so some objects with disks are observed during 
the non-accreting epochs \citep[see][]{Hartmann2005, Lada2006, sa2006, Scholz2007}.  
 
\subsection{Binarity}\label{bin}
Theoretical models of the evolution of T~Tauri disks in binary systems show that the individual circumstellar disks 
are truncated due to tidal effects. Once truncated, the two disks evolve independently, unless gap-clearing and 
circumbinary disk formation occur \citep{LA1996}. 

\begin{figure}[htp]
\begin{center}
\includegraphics[scale=.8]{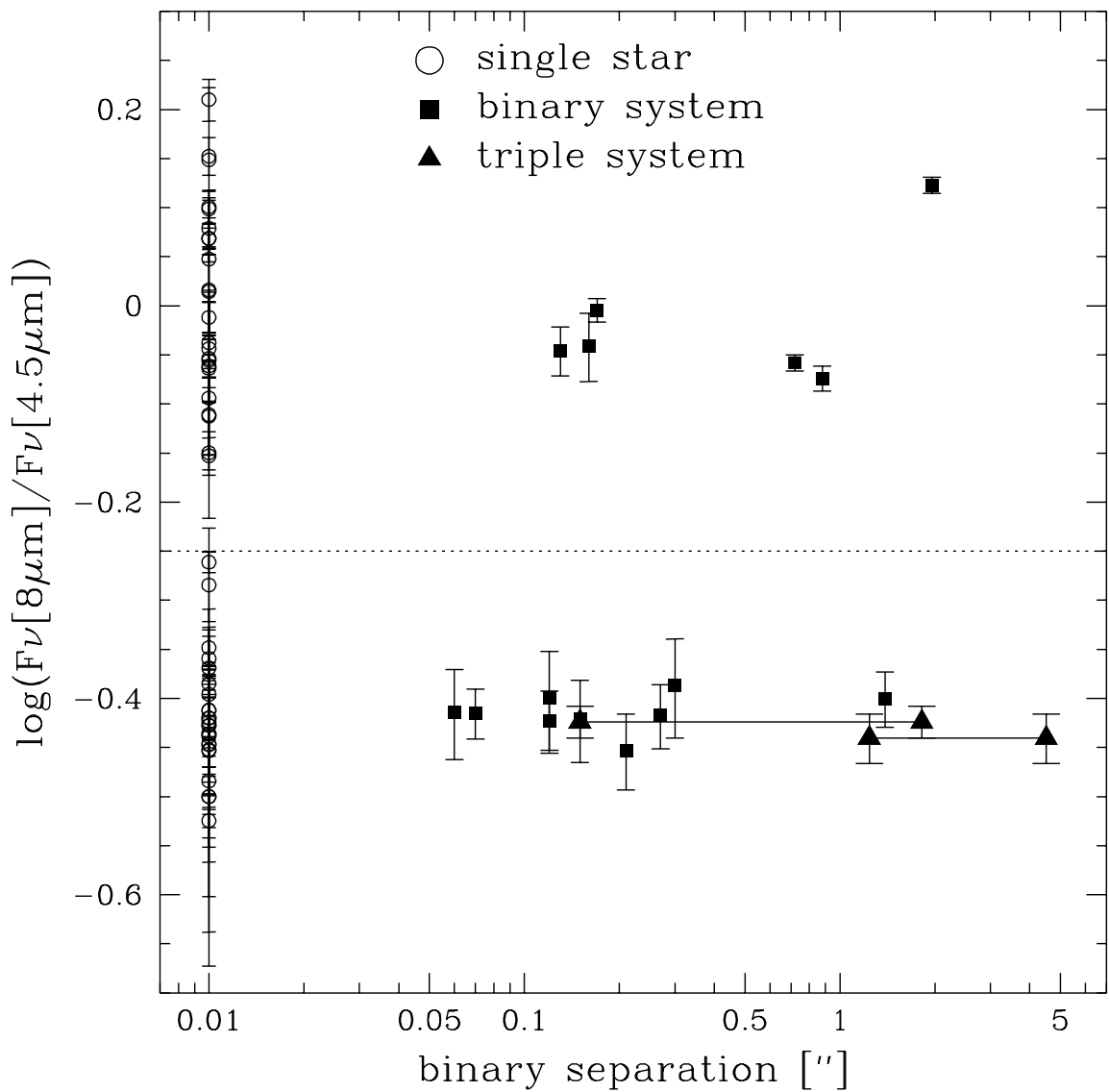}
\caption{8~$\mu$m excess, indicating presence of a disk, vs. binary separation for 65 Cha~I members (from 
Ahmic et al., in prep.). The \it{dotted line} \rm represents the threshold for 8~$\mu$m excess from a circumstellar disk, 
taken from the upper panel of Fig.~\ref{f3}. Two triple systems are presented with both separation values 
connected by \it{solid lines}\rm. Single objects are positioned at the binary separation of $0\farcs01$ for 
presentation purposes only.
\label{f8}}
\end{center}
\end{figure}

Of the Cha~I members in our sample, 65 have also been imaged with AO on the ESO VLT (Ahmic et al., in prep.). Among them, 
there are 15 binaries and two triple systems. In Fig.~\ref{f8}, we show the 8/4.5~$\mu$m flux ratio against 
projected binary separation for this sub-sample. Given the relatively poor angular resolution of IRAC and MIPS 
(\S~\ref{data}), these multiple systems are unresolved in the \it{Spitzer} \rm images; thus, 
we cannot determine which companion is responsible for the infrared excess. Nine of the binaries do not show infrared 
excess, while other six appear to harbor disks based on the 8/4.5~$\mu$m flux ratio. Neither of the two 
triples shows mid-infrared excess. 

The projected separations of the AO-resolved binaries range from $\sim10$~AU to $\sim300$~AU. Of the five binaries 
with $\lesssim20$~AU separations, four --including the two closest pairs-- exhibit low 8/4.5~$\mu$m flux ratios 
typical for diskless objects. This implies that either the disks have dissipated completely or a large inner hole 
is carved out by the companion; in the latter scenario, the disk would be circumbinary and undetectable with the 
near- and mid-IR data used in our analysis. The closest pair with detectable disk excess --Hn~13-- is separated by 
$\sim20$~AU. Theoretical modeling suggests that tidally truncated disks would have outer radii between 0.3--0.4 times 
the physical separation between the two stars for binaries with mass ratios close to one \citep{PP1977}. Thus, the 
disk(s) in the Hn~13 binary system would have an outer radius of $\sim8$~AU: emission from dust within this radius 
may explain the excess at 8 and 24~$\mu$m. 
  
The two triple systems detected, CHXR~28 and T39, do not exhibit 8~$\mu$m or 24~$\mu$m excess. Their projected 
separations are 24, $\sim290$~AU and $\sim200$, $\sim720$~AU, respectively. Again, the lack of mid-infrared excess 
may indicate faster disk dissipation due to the presence of companions or growth of large inner holes.

In a recent study of the $\simeq8$~Myr old $\eta$~Chamaeleontis star cluster, \citet{Bouwman2006} reported that 
disks clear out more rapidly in binary systems with projected separations $\lesssim20$~AU. Their analysis of 15 
K- and M-type stars, using their own \it{Spitzer} \rm   observations and AO-imaging by \citet{Brandeker2006}, shows that 
only one out of six binaries ($17^{+28}_{-14}\%$) retains a dusty disk while eight out of nine single stars 
($89^{+9}_{-21}\%$) harbor disks. Thus, they concluded that circumstellar disks and binarity are anti-correlated. 
We also find that the disk excess fraction is lower among multple systems ($35^{+15}_{-13}\%$) than among 
(apparently) single stars ($48\%\pm8\%$) in Cha~I, but the difference is less pronounced in this (younger) region. 
(The disk frequency for the total sample of 65 objects is $45\%\pm7\%$, a value slightly higher than that for 
multiple systems.) If we consider only those binaries with $\lesssim20$~AU separations, 4/5 do lack infrared 
excess. It may be that companions at (projected) distances $\gtrsim20$~AU do not have a strong effect 
on inner disk dissipation (at least not within $\simeq2$~Myr). 

\section{Summary}

We have presented a comprehensive view on the disks around stars and brown dwarfs in the 
nearby, young Cha~I dark cloud. In the following, we summarize our results:

\begin{enumerate}

\item{We used archival \it{Spitzer} \rm data to measure
mid-infrared fluxes in the wavelength range 3-24~$\mu$m for 81 known members of the Cha~I
star forming region. Our sample shows a clear bimodality in their \it{Spitzer} \rm   colours (flux
ratios 8/4.5~$\mu$m and 24/4.5~$\mu$m). For about half of the objects, the mid-infrared
colours are consistent with purely photospheric emission, while the other half shows substantial 
mid-infrared colour excess, interpreted as emission from a dusty disk. In both colours,
there is a clear gap between the two groups of objects.}

\item{The overall disk frequency in our sample is $52\%\pm6\%$ based on the 8/4.5~$\mu$m 
colour and $58^{+6}_{-7}\%$ based on the 24/4.5~$\mu$m colour. This value is consistent with the 
disk fractions found in similarly aged regions like IC~348 and Taurus \citep{Haish2001,Hartmann2005,Lada2006}. 
In contrast, the stellar disk frequency in the somewhat older star forming region Upper Scorpius (age $\sim5$~Myr) appears to 
be significantly lower ($\sim20$\%, \citealt{Carpenter2006}).}

\item{The disk fraction in our sample is, within the statistical uncertainties, not a function 
of spectral type: independent of the chosen binning, it ranges between 47\% and 55\% for 
spectral types between K3 and M8. This is in contrast to recent findings in other star forming
regions, where the disk frequency has been found to change significantly in the same spectral range
\citep{Lada2006,Carpenter2006,Scholz2007}. Environmental differences and
mass-dependent disk evolution might be important aspects to explain these findings.}

\item{There is no clear difference in the spatial distribution of stars with and without disks. 
The presence of bright stars in Cha~I does not seem to have any effect on the spatial distribution of dusty disks 
at distances grater than few arcmin. Although average positions of stars with disks within the two Cha~I cores are 
slightly displaced with respect to the average positions of disk-bearing stars, the difference is too small to be statistically important.}

\item{In our sample, there are no definitive transition disks, i.e. objects without excess at 
3--8~$\mu$m, but significant excess at longer wavelengths. These transition disks are often discussed
as objects caught in the process of inner disk clearing. However, we find two objects which 
have weak excess emission at 3--8~$\mu$m (and no colour excess at these wavelengths), but
substantial excess at 24~$\mu$m. These objects may be potential pre-cursors to transition disks. This 
type of SED is best seen in the late-K type star T35, which shows a rising SED at 
24~$\mu$m.}

\item{For 35 objects in our sample, we combined the \it{Spitzer} \rm   data with results from high-resolution 
optical spectroscopy. As a proxy for accretion, we use the H$\alpha$ 10\% width and adopt a 
threshold of 200\,km~s$^{-1}$ between accretors and non-accretors. Among the vast majority 
(29/35), there is a correlation between accretion signatures and infrared excess, consistent with the 
idea that the dust clearing of the inner disk coincides with a cessation of the gas accretion. However, 
six of the non-accretors still show signs of an inner dust disk. Thus, in some cases young objects can retain
their disks, after continuous accretion has stopped or at least dropped below measurable levels. 
}

\item{We use the results from a Cha~I imaging survey with high spatial resolution using the AO system NACO
at ESO/VLT to investigate the impact of multiplicity on the disk properties. AO images have been obtained for 
65 objects in our \it{Spitzer} \rm   sample, 17 of them have companions at separations between $0\farcs06$
and $5\arcsec$, corresponding to (projected) physical separations of 10--800~AU. The disk frequency among the 
objects with companions ($35^{+15}_{-13}\%$) is lower than (but still consistent with) the value for the total 
sample, which might 
hint that the presence of a companion contributes to faster disk dispersal. However, the existence of six
objects with disks and companions (at separations between 20--300~AU) suggests that the dispersion of the
inner disk is not strongly affected by companions at $\gtrsim20$~AU. For the closest binary with disk, the companion sets 
an upper limit for the outer disk radius of $\sim8$~AU. (Note that all multiple systems are unresolved
in \it{Spitzer} \rm   images, implying that we cannot determine which component is responsible for the mid-infrared excesss.)}

\end{enumerate}

\acknowledgments 
This research was supported by NSERC grants to RJ and MHvK, and an Early Researcher Award from the Province of 
Ontario to RJ.

\clearpage

\clearpage

\clearpage

\clearpage

\clearpage

\clearpage

\clearpage

\clearpage

\clearpage

\begin{deluxetable}{llcccccc}
\tablecolumns{8}
\tablewidth{0pc}
\tablecaption{Chamaeleon~I members with detectable infrared fluxes\label{tbl-1}}
\tabletypesize{\tiny}
\tablehead{
\colhead{ID}   &  \colhead{SpT\tablenotemark{a}} & \colhead{\it J-H\rm\tablenotemark{b}} &  
\colhead{3.6 $\mu$m\tablenotemark{c}} &  \colhead{4.5 $\mu$m\tablenotemark{c}} & 
\colhead{5.8 $\mu$m\tablenotemark{c}}&  \colhead{8 $\mu$m\tablenotemark{c}} &  \colhead{24 $\mu$m\tablenotemark{d}} \\
\cline{4-8}
\colhead{} & \colhead{} & \colhead{(mag)} &  \multicolumn{5}{c}{(mJy)}}
\startdata
ISO~28\dotfill&M5.5&0.89&9.1 $\pm$ 0.7&6.5 $\pm$ 0.7&4.6 $\pm$ 0.9&2.0 $\pm$ 0.5&0.6\tablenotemark{e}\\
Hn~2\dotfill&M5&0.90&37 $\pm$ 2&25 $\pm$ 1&17 $\pm$ 2& 8.9 $\pm$ 0.7&1.2 $\pm$ 0.6\\
CHXR~12\dotfill&M3.5&0.80&45 $\pm$ 2&30 $\pm$ 2&20 $\pm$ 2&10.9 $\pm$ 0.7&1.4 $\pm$ 0.6\\
ISO~52\dotfill&M4&0.81&29 $\pm$ 1&24 $\pm$ 1&17 $\pm$ 2&18.4 $\pm$ 0.9&40 $\pm$ 4\\
T16\dotfill&M3&1.21&\nodata&26 $\pm$ 1&\nodata&21 $\pm$ 1&\nodata\\
Hn~4\dotfill&M3.25&0.91&\nodata&39 $\pm$ 2&\nodata&13.8 $\pm$ 0.8&\nodata\\
CHXR~15\dotfill&M5.25&0.64&33 $\pm$ 2&24 $\pm$ 1&12 $\pm$ 2&9.7 $\pm$ 0.7&1.1 $\pm$ 0.5\\
Cam2-19\dotfill&M2.75&1.66&35 $\pm$ 2&23 $\pm$ 1&15 $\pm$ 1&9.6 $\pm$ 0.8&1.0 $\pm$ 0.5\\ 
CHXR~73\dotfill&M3.25&1.36&22 $\pm$ 1&14.5 $\pm$ 0.8&10 $\pm$ 2&5.1 $\pm$ 0.6&1.1 $\pm$ 0.5\\ 
Cha~H$\alpha$~12\dotfill&M6.5&0.71&8.6 $\pm$ 0.7&6.4 $\pm$ 0.7&6 $\pm$ 2&2.1 $\pm$ 0.5&0.9\tablenotemark{e}\\
ISO~79\dotfill& M5.25&1.64&8.3 $\pm$ 0.7&8.9 $\pm$ 0.8&4 $\pm$ 1&8.5 $\pm$ 0.7&16 $\pm$ 2 \\
Hn~5\dotfill&M4.5 &0.84&63 $\pm$ 2& 66 $\pm$ 2& 71 $\pm$ 2& 94 $\pm$ 2& 77 $\pm$ 5 \\
T22\dotfill&M3&1.03&65 $\pm$ 2&42 $\pm$ 2&32 $\pm$ 2&16.3 $\pm$ 0.9&1.9 $\pm$ 0.7\\
CHXR~20\dotfill&K6&0.99&113 $\pm$ 2&86 $\pm$ 2&81 $\pm$ 3&109 $\pm$ 2&99 $\pm$ 6\\
Ced~110-IRS~4\dotfill&\nodata&\nodata&19 $\pm$ 1&28 $\pm$ 2&30 $\pm$ 2&29 $\pm$ 1&\nodata\\
CHXR~74\dotfill&M3&0.93&30 $\pm$ 2&20 $\pm$ 1&12 $\pm$ 2&7.7 $\pm$ 0.7&1.0 $\pm$ 0.5\\
T23\dotfill&M1.5&0.78&32 $\pm$ 2&27 $\pm$ 1&20 $\pm$ 2&32 $\pm$ 2&47 $\pm$ 4\\
Ced~110-IRS~6\dotfill&\nodata&\nodata&197 $\pm$ 3& 314 $\pm$ 4& 356 $\pm$ 4& 394 $\pm$ 4&\nodata\\	    
ISO~91\dotfill&M2.5&2.32&27 $\pm$ 1&27 $\pm$ 1&23 $\pm$ 2&27 $\pm$ 1&45 $\pm$ 4\\
CHSM~9484\dotfill&M5.25&0.76&5.6 $\pm$ 0.6&2.7 $\pm$ 0.5&\nodata&3.0 $\pm$ 0.6&1.2 $\pm$ 0.6\\
T24\dotfill&M0.5&0.96&98 $\pm$ 2&79 $\pm$ 2&70 $\pm$ 2&70 $\pm$ 2&96 $\pm$ 5\\
CHXR~22E\dotfill&M3.5&1.32&40 $\pm$ 2&28 $\pm$ 1&21 $\pm$ 2&10.6 $\pm$ 0.8&11 $\pm$ 2\\
ISO 97\dotfill&M1&\nodata&42 $\pm$ 2&51 $\pm$ 2& 57 $\pm$ 2 & 64 $\pm$ 2&\nodata\\
Cha~H$\alpha$~1\dotfill&M7.75&0.67&6.7 $\pm$ 0.7&5.8 $\pm$ 0.7&4 $\pm$ 2&8.2 $\pm$ 0.7&\nodata\\
Cha~H$\alpha$~9\dotfill&M5.5&1.24&11.8 $\pm$ 0.8&12.0 $\pm$ 0.8&11 $\pm$ 2&11.0 $\pm$ 0.8&8 $\pm$ 2\\
B35\dotfill&M2&2.83&43 $\pm$ 2&51 $\pm$ 2&61 $\pm$ 2&82 $\pm$ 2&214 $\pm$ 8\\
CHXR~76\dotfill&M4.25&0.85&16.7 $\pm$ 0.9&11.6 $\pm$ 0.8&4 $\pm$ 2&4.2 $\pm$ 0.5&1.3 $\pm$ 0.6\\
CHXR~26\dotfill&M3.5&1.55&89 $\pm$ 2&63 $\pm$ 2&47 $\pm$ 2&25 $\pm$ 1&2.7 $\pm$ 0.8\\
Cha~H$\alpha$~7\dotfill&M7.75&0.71&5.5 $\pm$ 0.6&3.7 $\pm$ 0.5&4 $\pm$ 2&1.6 $\pm$ 0.4&\nodata\\
Cha~H$\alpha$~2\dotfill&M5.25&0.97&30 $\pm$ 1&26 $\pm$ 1&20 $\pm$ 2&24 $\pm$ 1&21 $\pm$ 3\\
CHXR~28\dotfill&K6&1.16&290 $\pm$ 4&186 $\pm$ 3&130 $\pm$ 3&70 $\pm$ 2&7 $\pm$ 2\\
T34\dotfill&M3.75&0.86&34 $\pm$ 2&23 $\pm$ 1&16 $\pm$ 2&8.3 $\pm$ 0.8&\nodata\\
Cha~H$\alpha$~13\dotfill&M5.5&0.73&22 $\pm$ 1&15.6 $\pm$ 0.9&9 $\pm$ 2&6.7 $\pm$ 0.7&\nodata\\
ISO~143\dotfill&M5&0.92&20 $\pm$ 1&20 $\pm$ 1&17 $\pm$ 2& 21 $\pm$ 1&19 $\pm$ 3\\
ISO~147\dotfill&M5.75&0.83&\nodata&5.0 $\pm$ 0.7&\nodata&4.8 $\pm$ 0.6&5 $\pm$ 1\\
T35\dotfill&K8&1.27&\nodata&100 $\pm$ 2&\nodata&45 $\pm$ 2&101 $\pm$ 6\\
Cha~H$\alpha$~6\dotfill&M5.75&0.78&19 $\pm$ 1&15.9 $\pm$ 0.9&11 $\pm$ 2&12.3 $\pm$ 0.8&12 $\pm$ 2\\
CHXR~33\dotfill&M0&0.90&67 $\pm$ 2&45 $\pm$ 2&36 $\pm$ 2&17.9 $\pm$ 0.9&1.8 $\pm$ 0.7\\
T37\dotfill&M5.25&0.73&15.1 $\pm$ 0.8&13.1 $\pm$ 0.9&11 $\pm$ 2&11.1 $\pm$ 0.7&9 $\pm$ 2\\
CHXR~78C\dotfill&M5.25&0.76&13.4 $\pm$ 0.8&10.2 $\pm$ 0.8&6 $\pm$ 2&4.0 $\pm$ 0.6&1.2\tablenotemark{e}\\
ISO~165\dotfill&M5.5&1.00&13.8 $\pm$ 0.9&13.4 $\pm$ 0.9&13 $\pm$ 2 &9.4 $\pm$ 0.7&9 $\pm$ 2\\
T39\dotfill&M2&0.78&124 $\pm$ 3&80 $\pm$ 2&53 $\pm$ 2&29 $\pm$ 1&3.3 $\pm$ 0.9\\
CHXR~35\dotfill&M4.75&0.64&17.6 $\pm$ 0.9&12.3 $\pm$ 0.8&8.0 $\pm$ 0.9&4.3 $\pm$ 0.5&1.2\tablenotemark{e}\\
CHXR~37\dotfill&K7&0.95&118 $\pm$ 2&77 $\pm$ 2&56 $\pm$ 2&29 $\pm$ 1&3.1 $\pm$ 0.9\\
CHXR~79\dotfill&M1.25&1.54&191 $\pm$ 3&191 $\pm$ 3& 163 $\pm$ 3&161 $\pm$ 3&230 $\pm$ 9\\
T40\dotfill&K6&1.23&406 $\pm$ 4&424 $\pm$ 4&370 $\pm$ 4&366 $\pm$ 4&272 $\pm$ 10\\
CHXR~40\dotfill&M1.25&0.84&87 $\pm$ 2&55 $\pm$ 2&39 $\pm$ 2&21 $\pm$ 1&2.2 $\pm$ 0.8\\
C7-1\dotfill&M5&1.15&43 $\pm$ 2 &35 $\pm$ 2&24 $\pm$ 2&18.1 $\pm$ 0.9&13 $\pm$ 2\\
B43\dotfill&M3.25&1.54&67 $\pm$ 2&69 $\pm$ 2&53 $\pm$ 2&49 $\pm$ 2&123 $\pm$ 6\\
ISO~209\dotfill&M1&2.0&\nodata&7.8 $\pm$ 0.7&\nodata&7.3 $\pm$ 0.7&8 $\pm$ 2\\
KG~102\dotfill&M5.5&0.83&10.0 $\pm$ 0.7&7.7 $\pm$ 0.7&6 $\pm$ 2&2.3 $\pm$ 0.5&1.2\tablenotemark{e}\\
ISO~217\dotfill&M6.25&0.99&15.3 $\pm$ 0.8&14.9 $\pm$ 0.9&13 $\pm$ 2&13.5 $\pm$ 0.8&12 $\pm$ 2\\
CHSM~15991\dotfill&M3&1.17&4.0 $\pm$ 0.6&5.9 $\pm$ 0.7&5 $\pm$ 2&7.5 $\pm$ 0.7&8 $\pm$ 2\\
ISO~220\dotfill&M5.75&1.28&7.5 $\pm$ 0.7&7.4 $\pm$ 0.7&5 $\pm$ 1&6.4 $\pm$ 0.7&7 $\pm$ 2\\
T43\dotfill&M2&1.29&109 $\pm$ 2&103 $\pm$ 2&107 $\pm$ 3&121 $\pm$ 3&185 $\pm$ 8\\
ISO~225\dotfill&M1.75&1.25&9.0 $\pm$ 0.7&12.0 $\pm$ 0.8&21 $\pm$ 2&22 $\pm$ 1&68 $\pm$ 5 \\
T45\dotfill&K8&1.07&458 $\pm$ 5&446 $\pm$ 4&393 $\pm$ 4&390 $\pm$ 4&303 $\pm$ 11\\
T46\dotfill&K8&0.95&235 $\pm$ 3&218 $\pm$ 3&218 $\pm$ 3&255 $\pm$ 3&236 $\pm$ 9\\
ISO~235\dotfill&M5.5&1.45&18.9 $\pm$ 0.9&17.1 $\pm$ 0.9&19 $\pm$ 2&17.7 $\pm$ 0.9&16 $\pm$ 2\\
CHSM~17173\dotfill&M8&0.63&4.3 $\pm$ 0.6&3.2 $\pm$ 0.6&7 $\pm$ 1&1.4 $\pm$ 0.5&0.6\tablenotemark{e}\\
Hn~12W\dotfill&M5.5&0.62&\nodata&14.8 $\pm$ 0.9&\nodata&5.5 $\pm$ 0.6&1.2\tablenotemark{e}\\
2MASS~11103481-7722053&M4&1.32&40 $\pm$ 2&30 $\pm$ 2&23 $\pm$ 2&11.3 $\pm$ 0.8&1.4 $\pm$ 0.6\\
ISO~250\dotfill&M4.75&1.36&26 $\pm$ 1&20 $\pm$ 1&13 $\pm$ 2&5.9 $\pm$ 0.7&1.0 $\pm$ 0.5\\
CHXR~47\dotfill&K3&1.05&241 $\pm$ 3& 201 $\pm$ 3&193 $\pm$ 3&199 $\pm$ 3&227 $\pm$ 9\\
ISO~252\dotfill&M6&0.97&7.0 $\pm$ 0.7&7.0 $\pm$ 0.7&5 $\pm$ 2&7.8 $\pm$ 0.7& 9$\pm$ 2\\
ISO~256\dotfill&M4.5&1.76&45 $\pm$ 2&48 $\pm$ 2&42 $\pm$ 2&50 $\pm$ 2&66 $\pm$ 5\\
Hn~13\dotfill&M5.75&0.74&56 $\pm$ 2&50 $\pm$ 2&48 $\pm$ 2&45 $\pm$ 2&54 $\pm$ 4\\
CHXR~48\dotfill&M2.5&0.78&45 $\pm$ 2&29 $\pm$ 1&19 $\pm$ 2&11.1 $\pm$ 0.7&1.3 $\pm$ 0.6\\
T49\dotfill&M2&0.96&\nodata&120 $\pm$ 3&\nodata&117 $\pm$ 2&\nodata\\
CHX~18N\dotfill&K6&0.83&\nodata&358 $\pm$ 4&\nodata&196 $\pm$ 3&\nodata\\
CHNR~49NE\dotfill&M0&0.67&\nodata&49 $\pm$ 2&\nodata&18.8 $\pm$ 0.9&\nodata\\
CHXR~84\dotfill&M5.5&0.66&19 $\pm$ 1&14.8 $\pm$ 0.8&12 $\pm$ 2&5.5 $\pm$ 0.6&1.2\tablenotemark{e}\\
ISO~282\dotfill&M4.75&1.04&13.2 $\pm$ 0.8&12.2 $\pm$ 0.8&10 $\pm$ 2&9.3 $\pm$ 0.7&9 $\pm$ 2\\
T50\dotfill&M5&0.78&52 $\pm$ 2&44 $\pm$ 2&35 $\pm$ 2&39 $\pm$ 2&50 $\pm$ 4\\
T51\dotfill&K3.5&0.75&377 $\pm$ 4&367 $\pm$ 4&344 $\pm$ 4&487 $\pm$ 5&242 $\pm$ 9\\
CHXR~55\dotfill&K4.5&0.66&60 $\pm$ 2&38 $\pm$ 2&24 $\pm$ 2&14.0 $\pm$ 0.8&1.8 $\pm$ 0.7\\
Hn~18\dotfill&M3.5&0.75&27 $\pm$ 1&24 $\pm$ 1&20 $\pm$ 2&19.1 $\pm$ 0.9&24 $\pm$ 3\\
CHXR~59\dotfill&M2.75&0.74&46 $\pm$ 2&31 $\pm$ 1&19 $\pm$ 2&11.6 $\pm$ 0.7&1.2 $\pm$ 0.6\\
CHXR~60\dotfill&M4.25&0.71&22 $\pm$ 1&17.1 $\pm$ 0.9&11 $\pm$2&5.4 $\pm$ 0.5&1.2\tablenotemark{e}\\
T55\dotfill&M4.5&0.60&20 $\pm$ 1&12.8 $\pm$ 0.9&13 $\pm$ 2&4.5 $\pm$ 0.6&1.2\tablenotemark{e}\\
CHXR~62\dotfill&M3.75&0.82&33 $\pm$ 2&23 $\pm$ 1&15 $\pm$ 2&9.3 $\pm$ 0.7&1.4 $\pm$ 0.6\\
\enddata
\tablenotetext{a}{\ Luhman 2004}
\tablenotetext{b}{\ 2MASS Point Source Catalog}
\tablenotetext{c}{\ IRAC data}
\tablenotetext{d}{\ MIPS data}
\tablenotetext{e}{\ upper limit}
\end{deluxetable}

\clearpage
\begin{deluxetable}{llcccc}
\tablecolumns{5}
\tablewidth{0pc}
\tablecaption{Chamaeleon~I members with accretion signatures measurements and multiplicity status\label{tbl-2}}
\tabletypesize{\tiny}
\tablehead{
\colhead{ID}   &  \colhead {SpT} &\colhead{H$\alpha$~10\%~width} & \colhead{Reference} &  
\colhead{Multiplicity} &  \colhead{Projected separation} \\ 
\colhead{} & \colhead {} &\colhead{(km~s$^{-1}$)} & \colhead{} & \colhead{}  & \colhead{(AU)}
}
\startdata
CHXR~12\dotfill&M3.5&110 $\pm$ 20& 1&single&\nodata \\
ISO~52\dotfill&M4&107 $\pm$ 8& 1&single&\nodata \\
Cha~H$\alpha$~12\dotfill&M6.5&107&2&single&\nodata \\
Hn~5\dotfill&M4.5&340 $\pm$ 20&1&single&\nodata \\
CHXR~20\dotfill&K6&70 $\pm$ 30&1&single&\nodata \\
CHXR~74\dotfill&M3&100 $\pm$ 20&1&single&\nodata \\
T23\dotfill&M1.5&247&2&single&\nodata \\
T24\dotfill&M0.5&430 $\pm$ 50&1&single&\nodata \\
Cha~H$\alpha$~1\dotfill&M7.75&151&2&single&\nodata \\
Cha~H$\alpha$~9\dotfill&M5.5&91&2&single&\nodata \\
CHXR~76\dotfill&M4.25&88 $\pm$ 7&1&single&\nodata \\
Cha~H$\alpha$~7\dotfill&M7.75&114 &2&single&\nodata \\
Cha~H$\alpha$~2\dotfill&M5.25&317&2&binary&$\sim25$\\
T34\dotfill&M3.75&90 $\pm$ 20&1&single&\nodata \\
Cha~H$\alpha$~13\dotfill&M5.5&122&1&single&\nodata\\
T35\dotfill&K8&450 $\pm$ 30 &1&single&\nodata\\
Cha~H$\alpha$~6\dotfill&M5.75&282&1&single&\nodata\\
CHXR~33\dotfill&M0&129 $\pm$2&1&single&\nodata\\
CHXR~37\dotfill&K7&140 $\pm$ 30&1&binary&$\sim11$\\
T40\dotfill&K6&350 $\pm$ 40&1&single&\nodata\\
CHXR~40\dotfill&M1.25&130 $\pm$ 20&1&binary&$\sim19$\\
ISO~217\dotfill&M6.25&334&2&single&\nodata\\
T43\dotfill&M2&170 $\pm$ 80&1&single&\nodata\\
T45\dotfill&K8&620 $\pm$ 30&1&binary&$\sim115$\\
T46\dotfill&K8&320 $\pm$ 60&1&single&\nodata\\
ISO~252\dotfill&M6&274&2&single&\nodata\\
CHXR~48\dotfill&M2.5&110 $\pm$ 10&1&single&\nodata\\
T49\dotfill&M2&280 $\pm$ 30&1&single&\nodata\\
CHX~18N\dotfill&K6&400 $\pm$ 40&1&single&\nodata\\
T50\dotfill&M5&250 $\pm$ 60&1&single&\nodata\\
T51\dotfill&K3.5&240 $\pm$ 30&1&single&\nodata\\
Hn~18\dotfill&M3.5&120 $\pm$ 40&1&single&\nodata\\
CHXR~59\dotfill&M2.75&130 $\pm$20&1&binary&$\sim24$\\
CHXR~60\dotfill&M4.25&100 $\pm$ 20&1&single&\nodata\\
CHXR~62\dotfill&M3.75&140 $\pm$ 20&1&binary&$\sim19$\\
\enddata
\tablerefs{(1) Nguyen et al., in prep.;(2)~\citealp{Mohanty2005,Muzerolle2005}}
\end{deluxetable}




\end{document}